\documentclass[amsmath,amssymb,twocolumn,nofootinbib]{revtex4}

\usepackage{latexsym}
\usepackage{amssymb}
\usepackage{amsfonts}
\usepackage{amsmath}
\usepackage[dvips]{graphicx}
\usepackage{bm}

\def\ber{\begin{eqnarray}}
\def\eer{\end{eqnarray}}
\def\beq{\begin{equation}}
\def\eeq{\end{equation}}

\begin{document}

\title{Gravito-magnetic gyroscope precession in Palatini $f(R)$  gravity}

\author{Matteo Luca Ruggiero}
\email{matteo.ruggiero@polito.it}
 \affiliation{Dipartimento di Fisica, Politecnico di Torino, Corso Duca degli Abruzzi 23, Torino, Italy\\
 INFN, Sezione di Torino, Via Pietro Giuria 1, Torino, Italy}

\date{\today}

\begin{abstract}
We study gravito-magnetic effects in the Palatini formalism of
$f(R)$ gravity. On using the Kerr-de Sitter metric, which is  a
solution of $f(R)$ field equations, we  calculate the impact of
$f(R)$ gravity on  the gravito-magnetic precession of an orbiting
gyroscope. We show that, even though an $f(R)$ contribution is
present in principle, its magnitude is negligibly small and far to
be detectable in the present (like GP-B) and foreseeable space
missions or observational tests around the Earth.
\end{abstract}

\maketitle

\section{Introduction} \label{sec:intro}

Among the theories that have been proposed to explain  the present
acceleration of the Universe \citep{Riess98,Perlmutter99,Bennet03}
without requiring the existence of a dark energy, $f(R)$ theories
of gravity received much attention in recent years. In these
theories the gravitational Lagrangian depends on an arbitrary
function $f$ of the scalar curvature $R$;  they are also referred
to as ``extended theories of gravity'', since they naturally
generalize, on a geometric ground, General Relativity (GR):
namely, when $f(R)=R$ the action reduces to the usual
Einstein-Hilbert action, and Einstein's theory is obtained.
Extended theories of gravity can be studied in different (non
equivalent) formalisms  (see \citet{Capofranc07} and references
therein): in order to obtain the field equations in the
\textit{metric formalism} the action is varied with respect to
metric tensor only; in the \textit{Palatini formalism} the action
is varied with respect to the metric and the affine connection,
which are supposed to be independent from one another. Actually,
$f(R)$ provide cosmologically viable models, where both the
inflation phase and the accelerated expansion are reproduced (see
\citet{Nojiri07} and references therein) and, furthermore, they
have been used to reproduce the rotation curves of galaxies
without need for dark matter \citep{Capo07,Martins2007}. It is
worthwhile noticing that,  because of the excellent agreement of
GR with Solar System and binary pulsar observations, every
modified theory of gravity should have the correct Newtonian and
post-Newtonian limits, in order to agree with GR tests (see e.g.
\citet{Will06}), and this is an important issue for $f(R)$ gravity
too (for a comprehensive  review of $f(R)$ theories, where this
and other issues are dealt with in details, we refer to the recent
paper by \citet{sotfar08}).

In this paper we are concerned with gravito-magnetic (GM) effects
in $f(R)$ theories of gravity. GM effects are post-Newtonian
effects originated by the rotation of the sources of the
gravitational field: this gives raise to the presence of
off-diagonal terms in the metric tensor, which are responsible for
the dragging of the inertial frames. These effects are expected in
GR, but are generally very small and, hence, very difficult to
detect \citep{mashh1,ruggiero02,mashhoon03}. In recent years,
there have been some attempts to measure these effects (see e.g.
\citet{ciufolini04}, \citet{iorio05} and references therein);  in
April 2004 Gravity Probe B was launched to accurately measure the
frame dragging (and the geodetic precession) of an orbiting
gyroscope: the final results are going to be published
\citep{gpb}.

Here,  working on an exact solution of the field equations in the Palatini formalism (GM effects and other
Post-Newtonian effects were obtained in metric $f(R)$ gravity by
\citet{clifton08}), we want to evaluate the impact of $f(R)$
theories on GM effects, in order to see if there are corrections
to the GR predictions that can be detected (at least in principle)
by GP-B or other foreseeable experiments around the Earth.

\section{Vacuum field equations of Palatini $f(R)$ gravity} \label{sec:theof}

The equations of motion of  $f(R)$ extended theories of gravity
can be obtained starting from the action:

\begin{equation}
A=A_{\mathrm{grav}}+A_{\mathrm{mat}}=\int [ \sqrt{g} f (R)+2\chi
L_{\mathrm{mat}} (\psi, \nabla \psi) ]  \; d^{4}x.
\label{eq:actionf(R)}
\end{equation}
The gravitational part of the Lagrangian is represented by a
function $f (R)$ of  the scalar curvature $R$. The total
Lagrangian contains also a first order matter part
$L_{\mathrm{mat}}$ functionally depending on matter fields $\Psi$,
together with their first derivatives, equipped with a
gravitational coupling constant $\chi=\frac{8\pi G}{c^4}$. In the
Palatini formalism the metric $g$ and the affine connection
$\Gamma$ are supposed to be independent, so that the scalar
curvature $R$ has to be intended as $R\equiv R( g,\Gamma)
=g^{\alpha\beta}R_{\alpha \beta}(\Gamma )$, where $R_{\mu \nu
}(\Gamma )$ is the Ricci-like tensor of the connection $\Gamma$.

By independent variations with respect to the metric $g$ and the
connection $\Gamma$, we obtain the following equations of motion:
\begin{eqnarray}
f^{\prime }(R) R_{(\mu\nu)}(\Gamma)-\frac{1}{2} f(R)  g_{\mu \nu
}&=&\chi T_{\mu \nu },  \label{ffv1}\\
\nabla _{\alpha }^{\Gamma }[ \sqrt{g} f^\prime (R) g^{\mu \nu
}]&=&0, \label{ffv2}
\end{eqnarray}
where $f^{\prime }(R)=df(R)/dR$, $T_{\mu\nu}$ is the matter source stress-energy tensor and
$\nabla^{\Gamma}$ means covariant derivative with respect to the
connection $\Gamma$ \citep{Capofranc07,sotfar08}.

The equation of motion (\ref{ffv1}) can be supplemented by the
scalar-valued equation obtained by taking the contraction of
(\ref{ffv1}) with the metric tensor:
\begin{equation}
f^{\prime} (R) R-2 f(R)= \chi T,  \label{ss}
\end{equation}
where $T$ is the trace of the energy-momentum tensor. Equation
(\ref{ss}) is an algebraic equation for the scalar curvature $R$:
it is called the
\textit{structural equation} and it controls the solutions of equation (\ref{ffv1}).\\

We are interested into solutions of the field equation in vacuum,
in particular outside a rotating source of matter: so the field
equations become
\begin{eqnarray}
f^{\prime }(R) R_{(\mu\nu)}(\Gamma)-\frac{1}{2} f(R)  g_{\mu \nu
}&=&0,  \label{ffv1vac}\\
\nabla _{\alpha }^{\Gamma }[ \sqrt{g} f^\prime (R) g^{\mu \nu
}]&=&0, \label{ffv2vac}
\end{eqnarray}
and they are, again, supplemented by the scalar equation
\begin{equation}
f^{\prime} (R) R-2 f(R)=0 . \label{ssvac}
\end{equation}

The trace equation (\ref{ssvac}) is an algebraic equation for $R$
which admits constant solutions $R=c_{i}$. Then, it is possible to
show that, under suitable conditions (see
\citet{FFVa},\citet{allemandi05}), the field equations
(\ref{ffv1vac},\ref{ffv2vac}) reduce to
\begin{equation}
R_{\mu \nu }\left( g\right)=k
 g_{\mu \nu },
  \label{eq:genein1}
\end{equation}
with $k=c_i/4$, which are identical to GR equations with a
cosmological constant $\Lambda$: in practice, it is $\Lambda=k$ in
our notation. Indeed, we may say that $k$ is a measure of the
non-linearity of the theory (if $f(R)=R$, eq. (\ref{ssvac}) has only
the solution $R=0 \rightarrow k=0$).

Consequently, the vacuum solutions of GR with a cosmological
constant can be used in  Palatini $f(R)$ gravity: the role of the
$f(R)$ function is determining the solutions of the structural
equation (\ref{ssvac}). It is useful to point out that, for a
given $f(R)$ function, in vacuum case the solutions of the field
equations of Palatini $f(R)$ gravity are a subset of the solutions
of the field equations of metric $f(R)$ gravity (see e.g.
\citet{magnano}). So every solution of eqs. (\ref{eq:genein1}) is
also a solution of the field equations of metric $f(R)$ gravity
\textit{with constant scalar curvature $R$}.\\

The Kerr-de Sitter  metric, which is an exact solution of the field equations in the form (\ref{eq:genein1}),
describes a rotating black-hole
in a space-time with a cosmological constant
\citep{kerr03,demianski73,carter73} and can be used to investigate
GM effects in extended theories of gravity.

The Kerr-de Sitter metric in the standard Boyer-Lindquist
coordinates $x^{\mu} = (t, r, \theta, \phi)$ has the form
\footnote{The space-time metric has signature $(-1,1,1,1)$,  we
use geometrized units such that $G=c=1$, greek indices run from 0
to 3, and latin ones run from 1 to 3, boldface letters like
$\bm{x}$ refers to three-vectors.}
\begin{widetext}
\begin{eqnarray} ds^2=&-& \left[ 1 - \frac{2Mr}{\Sigma} - \frac{k }{3} (r^2 + a^2\; {\rm
sin}^2\theta)\right] dt^2  - 2a\left[\frac{2Mr}{\Sigma} + \frac{k }{3} (r^2 +
a^2)\right]\;{\rm sin}^2\theta dtd\phi\nonumber\cr\\ &+& \frac{\Sigma}{\Delta} dr^2 +
\frac{\Sigma}{\chi} d\theta^2 + \left[ \frac{2Mr}{\Sigma} a^2{\rm sin}^2\theta + (1+
\frac{k }{3} a^2) (r^2 + a^2)\right] \; {\rm sin}^2\theta d\phi^2\;\;, \label{eq:kdsmetric1}
\end{eqnarray}
\end{widetext}
where
\begin{equation}
\Sigma = r^2+ a^2\cos^2\theta\;\;,\;\;\chi = 1 + \frac{k }{3} a^2\cos^2\theta\;\;, \label{eq:kdsdef1}
\end{equation}

\begin{equation}
\Delta = r^2 - 2Mr + a^2 - \frac{k }{3} r^2(r^2 + a^2)\;\;. \label{eq:kdsdef2}
\end{equation}

The mass of the source is $M$, while $J = Ma$ is its angular
momentum (which is perpendicular to the $\theta=\pi/2$ plane).
When $k=0$  the Kerr-de Sitter metric $(\ref{eq:kdsmetric1})$
reduces to the Kerr metric. Other limiting cases can be checked:
for instance, when $a=0$, we obtain the Schwarzschild-de Sitter
solution, and when $M=a=0$ we have the de Sitter space-time.

It is interesting to consider the ``weak-field'' approximation of
the metric (\ref{eq:kdsmetric1}): in other words, we expand it up
to linear terms in $M/r,Ma/r^{2},kr^{2},kar,kMr$. What we get is
\begin{widetext}
\begin{equation}
ds^{2}=-\left(1-\frac{2M}{r}-\frac{k}{3}r^{2}
\right)dt^{2}+\left(1+ \frac{2M}{r}-\frac{k}{3}r^{2}
\right)dr^{2}+r^{2}d\theta^{2}+r^{2}\sin^{2}\theta^{2}d\phi^{2}-2a
\left(\frac{2M}{r}+\frac{k}{3}r^{2} \right)\sin^{2}\theta d\phi
dt. \label{eq:kdsweak1}
\end{equation}
\end{widetext}
We notice that on using Boyer-Lindquist coordinates the weak field metric (\ref{eq:kdsweak1}) does not contain terms
in the form $kMr$. However, these  coordinates are not directly related to physical lengths. 
We can express the metric (\ref{eq:kdsweak1}) in the more familiar
isotropic coordinates, which is necessary  to properly deal with
gravito-magnetic effects \citep{mashhoon03}. To this end we
introduce a new radial coordinate $\rho$

\begin{equation}
\rho = r \left(1-\frac{M}{r}-\frac{k r^2}{12} \right)
\label{eq:iso6},
\end{equation}
and, up to the required approximation order, the metric turns out
to be:
\begin{widetext}
\begin{align}
ds^2=&-\left(1-\frac{2M}{\rho}-\frac{k }{3}\rho^2\right)dt^2 +
\left(1+\frac{2M}{\rho} -\frac{k }{6}\rho^2
\right)\left(d\rho^2+\rho^2 d\theta^2+\rho^2 \sin^2 \theta d\phi^2\right)+ \notag \\
& -2a\left(\frac{2M}{\rho}+\frac{k }{3}\rho^2+\frac{5}{6} M k \rho
\right)\sin^2  \theta d\phi dt. \label{eq:kdsweak2}
\end{align}
\end{widetext}
On using these coordinates we see that terms in the form $kM\rho$  are  present:
as a consequence, their absence in the metric (\ref{eq:kdsweak2}) was due to the use of Boyer-Lindquist coordinates.

By inspection of the metric (\ref{eq:kdsweak2}) we see that for $a=0$, we obtain the weak field limit of
the Schwarzschild-de Sitter solution, for $k=0$ we obtain the weak
field limit of the Kerr solution.

\section{Gravito-magnetic field effects in Palatini $f(R)$ gravity} \label{sec:gmfr}

According to what we have seen, the terms containing $k$ in the
metric (\ref{eq:kdsmetric1}) are due to the non linearity of the
gravity Lagrangian in the framework of Palatini $f(R)$ gravity, or
to the presence of a cosmological constant in the framework of GR.
Some effects of these terms have been already investigated in the
literature. For instance  \citet{kerr03}, \citet{sereno06} showed
that, due to the cosmological constant,  the mean motion for
circular geodesics is modified according to \beq \omega_{k}=
\frac{M}{\rho^{3}}-\frac{k}{3}; \label{eq:kepmod1}\eeq and that
the precession of pericenter (in the $\theta=\pi/2$ plane) gains
an additional contribution: \beq \Delta \phi_{k}= \frac{\pi k
A^{3}}{M}\sqrt{1-e^{2}}, \label{eq:perimod1}\eeq where $A$ is the
semi-major axis of the (unperturbed) orbit, and $e$ is its eccentricity. The
effects of the cosmological constant on gravitational lensing was
recently studied by \citet{rindler07},\citet{sereno08} and by
\citet{ruggiero07} in connection with Palatini $f(R)$ gravity.
Actually, the effects (\ref{eq:kepmod1},\ref{eq:perimod1}) of the
$k$-term (whose expected order of magnitude is comparable to the
cosmological constant $k \simeq \Lambda \simeq 10^{-52} \
m^{-2}$), though present in principle, are too small to be
detected.

Here, we would like to study the GM precession of an orbiting
gyroscope,  focusing on the corrections due to the $k$-term.

We remember that such an orbiting gyroscope undergoes also a geodetic precession with
an angular frequency
(averaged over a revolution)  that can be written in
the form \beq
\bm{\Omega}_{S-O}=\bm{\Omega}_{S-O}^{M}+\bm{\Omega}_{S-O}^{k}
\label{eq:precpn3bis} \eeq where \beq \bm{\Omega}_{S-O}^{M}= \frac
3 2 \frac{M}{A^{3}}\frac{\bm{L}}{\left(1-e^{2}\right)^{3/2}}
\label{eq:precpn3bis1} \eeq is the classical GR geodetic (or de
Sitter) precession, where $\bm{L}$ is the specific orbit angular
momentum of the gyroscope, moving along the orbit with  semi-major
axis $A$ and eccentricity $e$, and \beq
\bm{\Omega}_{S-O}^{k}=-\frac 1 2 k \bm {L} \label{eq:precpn3bis2}
\eeq is the correction due to the $k$-term (see e.g.
\citet{sereno06}).

The angular velocity $\bm{\Omega}_{S-O}$ can be referred to as a
spin-orbit term, because it is due to the coupling of the
gyroscope's spin $\bm{S}$ with the orbit angular momentum
$\bm{L}$.

As for the GM precession, it can be calculated, using the standard approach (see
e.g. \citet{ciufo95}), starting from the GM potential 
\beq  g_{0\phi}=-2\frac{Ma}{\rho}\sin^2 \theta -\frac{ka
}{3}\rho^2\sin^2 \theta - \frac{5}{6} M a k \rho \sin^2 \theta
\label{gempot1}\eeq
of the metric (\ref{eq:kdsweak2}).  The first term in eq. (\ref{gempot1}) is the ``classical'' GM potential arising in GR,
while the other two terms are proportional to $k$. 
As a consequence, we can write the angular frequency of GM precession in the form

 \beq \bm{\Omega}_{S-S}=
\bm{\Omega}_{S-S}^{J}+\bm{\Omega}_{S-S}^{ka}+\bm{\Omega}_{S-S}^{Jk}
\label{eq:precpn4bis} \eeq where \beq \bm{\Omega}_{S-S}^{J} =  -
\left[\frac{\bm{J}}{|\bm{\rho}|^3}-3\frac{\left(\bm{J}\cdot
\bm{\rho} \right)\bm{\rho}}{|\bm{\rho}|^5}\right],
\label{eq:prec2}\eeq is the GR gravito-magnetic precession (see
e.g. \citet{ciufo95}), and \beq \bm{\Omega}_{S-S}^{ka} =
\frac{ka}{3} \hat{\bm{J}},  \label{eq:prec3} \eeq \beq
\bm{\Omega}_{S-S}^{Jk} =  \frac{5Jk}{12\rho} \left[ \hat{\bm{J}}+
\left(\hat{\bm{J}} \cdot \hat{\bm{\rho}} \right)\hat{\bm{\rho}}
\right]. \label{eq:prec33} \eeq are  new contributions due to the
presence of the $k$-term.

The angular velocity $\bm{\Omega}_{S-S}$ can be referred to as a
spin-spin term, because it is due to the coupling of the
gyroscope's spin $\bm{S}$ with the  spin angular momentum $\bm{J}$ of
the source of the gravitational field. We point out that all
contributions in (\ref{eq:precpn4bis}) do not depend on the
velocity of the gyroscope; furthermore, the term
$\bm{\Omega}_{S-S}^{ka}$ is a constant contribution over the whole
orbit.

It is interesting now evaluate the magnitude of the new
contributions (\ref{eq:prec3},\ref{eq:prec33}). To this end, it is
useful to remember that  the GP-B mission is expected to measure
the precession (geodetic plus frame-dragging) of the orbiting
gyroscope with an accuracy of 0.1 milliarcseconds/year, which is a
very hard task, as the long story of this mission teaches
\citep{gpb}. Taking into account that the geodetic precession has
a magnitude of about 6.6 arcseconds/year, and the frame-dragging
effect of 0.041 arcseconds/year, to give an idea of the magnitude
of the effects of the non linearity of the gravity Lagrangian, we
may calculate the ratio between the two angular frequencies
(\ref{eq:prec3}) and (\ref{eq:prec33}) and the GR one
(\ref{eq:prec2}), at a distance $R=|\bm{x}|=\rho$ from the source
\beq \frac{{\Omega}_{S-S}^{ka}}{{\Omega}_{S-S}^J} \simeq \frac{k
R^3}{M}, \label{eq:prec4} \eeq

\beq \frac{{\Omega}_{S-S}^{Jk}}{{\Omega}_{S-S}^J}  \simeq kR^2.
\label{eq:prec44}  \eeq

Using $k=10^{-52} m^{-2}$, i.e. the current estimate of the
cosmological constant, $M$ equal to the Earth mass,   $R \simeq
650 \ Km$, i.e. order of magnitude of the GP-B orbit, we get \beq
\frac{{\Omega}_{S-S}^{ka}}{{\Omega}_{S-S}^J} \simeq 10^{-28}.
\label{eq:prec5} \eeq

\beq \frac{{\Omega}_{S-S}^{Jk}}{{\Omega}_{S-S}^J} \simeq 10^{-39}.
\label{eq:prec55} \eeq

Accordingly,  we may conclude that the impact of $f(R)$ gravity on
GM gyroscope precession is very small, and completely negligible
for a mission like GP-B. For similar reasons, we can say that  $f(R)$ gravity  is not relevant for
other experiments aimed at the measurement of the gravito-magnetic field of the Earth, such as those performed in the past
with LAGEOS satellites (see \citet{ciufolini04}), or those that are planned in the next months such as 
LARES \citep{ciufo}.

On the other hand , if GP-B will confirm
the GR predictions for the orbiting gyroscope precession and no
additional terms will be seen, from the expected accuracy of 0.1
milliarcseconds/year, we might deduce an estimate for the $k$-term
in (\ref{eq:prec3}): $|k|\leq 10^{-26}m^{-2}$, which is
considerably greater than the current best estimates of the
cosmological constant.

\section{Conclusions} \label{sec:conc}

We have studied gravito-magnetic effects in the framework of
$f(R)$ gravity. Namely, thanks to the analogy, in the Palatini
formalism, between general relativistic vacuum field equations
with cosmological constant and vacuum $f(R)$ field equations, we
have considered the Kerr-de Sitter metric as a solution of $f(R)$
field equations. Since this metric describes a rotating
black-hole, it is suitable to evaluate the gravito-magnetic
effects. In particular, we have considered the weak-field
approximation of the Kerr-de Sitter metric (which, as far as we
know, has never been studied before) and then we have calculated
the contribution to the gravito-magnetic precession of an orbiting
gyroscope due to the non linearity of the gravity Lagrangian. We
have shown that, though present in principle, this contribution is
very small, and far to be detectable by a mission like GP-B and,
probably, also by other foreseeable tests around the Earth. This
confirms that the non-linearities appearing in $f(R)$ become
important on length scales much larger than the Solar System (e.g.
on the cosmological scale) and  their effects on local physics are
probably negligible.

\section*{ACKNOWLEDGMENTS}
The author would like to thank Prof. B. Mashhoon and Dr. M. Sereno for useful
discussions. The author
acknowledges financial support from the Italian Ministry of
University and Research (MIUR) under the national program ``Cofin
2005'' - \textit{La pulsar doppia e oltre: verso una nuova era
della ricerca sulle
pulsar}.\\


\begin{thebibliography}{200}
\bibitem[Riess et al.(1998)]{Riess98}  Riess, A.G. et al., Observational
Evidence from Supernovae for an Accelerating Universe and a
Cosmological Constant, {\it Astron. J.}, {\bf 116}, 1009-1038,
1998

\bibitem[Perlmutter et al.(1999)]{Perlmutter99} Perlmutter, S., et al., Measurements of Omega and Lambda from 42 High-Redshift
Supernovae, \textit{Astrophys. J.}, \textbf{517}, 565-586, 1999

\bibitem[Bennet et al.(2003)]{Bennet03} Bennet, C.L., et al., First-Year Wilkinson Microwave Anisotropy Probe (WMAP) Observations: Preliminary Maps and Basic
Results, \textit{Astrophys. J. Suppl.} \textbf{148}, 1-27, 2003

\bibitem[Capozziello and Francaviglia(2007)]{Capofranc07}
Capozziello,  S., and  Francaviglia,  M.,  Extended Theories of
Gravity and their Cosmological and Astrophysical Applications,
arXiv:0706.1146 [astro-ph], 2007

\bibitem[Nojiri and Odintsov(2007)]{Nojiri07} Nojiri, S.,
Odintsov, S.D., Modified $f(R)$ gravity unifying $R^m$ inflation
with $\Lambda CDM$ epoch, arXiv:0710.1738 [astro-ph], 2007

\bibitem[Capozziello et al.(2007a)]{Capo07}
Capozziello, S., Cardone, V.F., and Troisi, A., Low surface
brightness galaxies rotation curves in the low energy limit of
$R^n$ gravity: no need for dark matter?, {\it Mon. Not. Roy.
Astron. Soc.}, {\bf 375}, 1423-1440, 2007a

\bibitem[F. Martins and Salucci(2007)]{Martins2007}
  F.~Martins, C.,  and Salucci, P., Analysis of rotation curves in the framework of $R^n$ gravity,
 \textit{Mon.\ Not.\ Roy.\ Astron.\ Soc.\ }, {\bf 381}, 1103-1108, 2007

\bibitem[Will(2006)]{Will06}  Will, C.M.,The Confrontation between
General Relativity and Experiment,  {\it Living Rev. Relativity},
{\bf 9}, 2006, http://www.livingreviews.org/lrr-2006-3

\bibitem[Sotiriou and Faraoni(2008)]{sotfar08} Sotiriou, T., Faraoni, V., $f(R)$ Theories of Gravity,
arXiv:0805.1726 [gr-qc], 2008


\bibitem[Mashhoon et al.(2001)]{mashh1} Mashhoon, B., Gronwald, F., Lichtenegger, H.I.M.,
Gravitomagnetism and the Clock Effect,  \textit{ Lect. Notes
Phys.}  \textbf{562},  83-108, 2001

\bibitem[Ruggiero and Tartaglia(2002)]{ruggiero02} Ruggiero, M.L., Tartaglia,
A., Gravitomagnetic Effects,
 \textit{Il Nuovo Cimento B}  \textbf{117}, 743-768, 2002

\bibitem[Mashhoon (2007)]{mashhoon03} Mashhoon, B., Gravitoelectromagnetism: A Brief Review,
in \textit{The Measurement of Gravitomagnetism: A Challenging
Enterprise}, Iorio, L., Editor, Nova Science, New York,
arXiv:gr-qc/0311030, 2007



\bibitem[Ciufolini and Pavlis(2004)]{ciufolini04}
Ciufolini, I.,  and Pavlis, E.C.,  A confirmation of the general relativistic prediction of the LenseÐThirring effect,  \textit{Nature} \textbf{431}, 958, 2004

\bibitem[Iorio(2005)]{iorio05} Iorio, L., On the reliability of the so-far performed tests
for measuring the Lense–Thirring effect with the LAGEOS
satellites, \textit{New Astronomy} \textbf{10}, 603-615, 2005

\bibitem[Everitt et al.(2001)]{gpb}
C.W.F., Everitt et al., Gravity Probe B: Countdown to Launch,
\textit{ Lect. Notes Phys.}  \textbf{562}, 52-82, 2001; see also
the web site \verb|http://einstein.stanford.edu|

\bibitem[Clifton(2008)]{clifton08} Clifton, T., The Parameterised Post-Newtonian Limit
of Fourth-Order Thoeries of Gravity, \textit{Phys. Rev. D}
\textbf{77}, 024041, 2008


\bibitem[Ferraris et al.(1993)]{FFVa} Ferraris, M., Francaviglia, M., Volovich, I.,
The Universality of Einstein Equations,  \textit{Nuovo Cim. B}
\textbf{108}, 1313, 1993

\bibitem[Allemandi et al.(2005)]{allemandi05} Allemandi, G., Francaviglia, M.,
Ruggiero, M.L., Tartaglia, A., Post-Newtonian Parameters from
Alternative Theories of Gravity, \textit{Gen. Rel. Grav.}
\textbf{37}, 1891-1904, 2005

\bibitem[Magnano (1994)]{magnano} Magnano G., Kerr, A.W., Are there metric theories of gravity other than general relativity?,
[arXiv:gr-qc/9511027], 1994

\bibitem[Kerr et al.(2003)]{kerr03} Kerr, A.W., Hauck, J.C., Mashhoon
B., Standard clocks, orbital precession and the cosmological
constant, \textit{Class. Quantum Grav.} \textbf{20}, 2727–2736,
2003

\bibitem[Demianski(1973)]{demianski73} Demianski M., Some New Solutions of the Einstein Equations of
Astrophysical Interest, \textit{Acta Astronomica} \textbf{23},
197-232, 1973

\bibitem[Carter(1973)]{carter73} Carter, B., Black hole equilibrium states, in \textit{Black Holes (Les Houches
1972)}, Gordon and Breach, London, 1973

\bibitem[Sereno and Jetzer(2006)]{sereno06} Sereno, M., Jetzer, Solar and stellar system tests of the cosmological constant,  \textit{Phys. Rev. D} \textbf{73},
063004, 2006

\bibitem[Rindler and Ishak(2007)]{rindler07} Rindler, W., Ishak, M., The Contribution of the Cosmological Constant to the Relativistic Bending of Light Revisited,  \textit{Phys. Rev. D} \textbf{76},
043006, 2007

\bibitem[Sereno(2008)]{sereno08} Sereno, M., On the influence of the cosmological constant on gravitational lensing in small systems.,  \textit{Phys. Rev. D} \textbf{77},
043004, 2008

\bibitem[Ruggiero(2007)]{ruggiero07} Ruggiero, M.L., Gravitational Lensing and f(R) theories in the Palatini approach,  \textit{Gen. Rel. Grav.} to appear, arXiv:0712.3218[astro-ph], 2007

\bibitem[Misner,Thorne,Wheeler(1973)]{mtw} Misner, C.W., Thorne, K.S., Wheeler, J.A., \textit{Gravitation}, W. H. Freeman and Company, San
Francisco, 1973

\bibitem[Ciufolini and Wheeler(1995)]{ciufo95} Ciufolini, I., Wheeler J.A., \textit{Gravitation and
Inertia}, Princeton University Press, Princeton, 1995

\bibitem[Ciufolini(2004)]{ciufo} Ciufolini, I., LARES/WEBER-SAT, frame-dragging and fundamental physics, arXiv:gr-qc/0412001, 2004


\end{thebibliography}
\end{document}